\documentclass[a4paper]{article}
\usepackage[normalem]{ulem}
\usepackage{latexsym,amssymb,amsmath,amssymb,xspace,theorem}
\usepackage{epic,eepic}

\theoremstyle{plain} \theoremheaderfont{\scshape}

\newtheorem{Thm}{\bf Theorem}
\newtheorem{Lem}[Thm]{\bf Lemma}

\newtheorem{Prop}[Thm]{\bf Proposition}
\newtheorem{Cor}[Thm]{ \bf Corollary}
{\theorembodyfont{\rmfamily}

 \newtheorem{Problem}[Thm]{\bf Problem}
}

\newenvironment{Prf}{{\bf \noindent Proof } }{\hfill$\square$\\}

\newcommand{\ignore}[1]{}

\newcommand{\cqfd}{\unskip\kern 6pt\penalty 500
\raise -2pt\hbox{\vrule\vbox to 10pt{\hrule width 4pt
\vfill\hrule}\vrule}\par}

\newcommand{\dm}{$\delta$-minimum edge-colouring \xspace}
\newcommand{\dms}{$\delta$-minimum edge-colourings \xspace}

\input{epsf.sty}
\begin{document}

\title{On parsimonious edge-colouring of graphs with maximum degree three}

\author{J.-L. Fouquet and J.-M. Vanherpe\\
L.I.F.O., Facult\'e des Sciences, B.P. 6759 \\ Universit\'e d'Orl\'eans, 45067 Orl\'eans Cedex 2, FR
}
\date{}
\maketitle

\begin{abstract}
In a graph $G$ of maximum degree $\Delta$ let $\gamma$ denote the
largest fraction of edges that can be $\Delta$ edge-coloured. Albertson and Haas showed  that $\gamma \geq \frac{13}{15}$ when $G$
is cubic . We show here that this result can be
extended to graphs with maximum degree $3$ with the exception of a
graph on $5$ vertices. Moreover, there are exactly two  graphs with
maximum degree $3$ (one being obviously  the Petersen graph) for
which $\gamma = \frac{13}{15}$. This extends a result given by Steffen. These results are obtained by using structural properties of the so called $\delta$-minimum edge
colourings for graphs with maximum degree $3$.

\end{abstract}



\section{Introduction}
Throughout this paper, we shall be concerned with connected graphs
with maximum degree $3$. We know by Vizing's  theorem \cite{Viz}
that these graphs can be edge-coloured with $4$ colours. Let $\phi :\
E(G) \rightarrow \{\alpha,\beta,\gamma,\delta\}$ be a proper
edge-colouring of $G$. It is often of interest to try to use one
colour (say $\delta$) as few as possible. When it is optimal following this constraint, we shall say that such a parsimonious edge-colouring $\phi$ is
$\delta-minimum$.  In \cite{Fou} we gave without proof
(in French, see \cite{FouVan10a} for a translation) results on $\delta-minimum$
edge-colourings of cubic graphs. Some of them have been obtained
later and independently by Steffen \cite{Ste,Ste04}. Some other results which were not stated formally in \cite{FouPhD} are needed here. 
The purpose of Section \ref{section:tools} is to give those results as structural properties of \dms as well as others which will be useful in Section \ref{section:Apps}.

An edge colouring of $G$ using colours $\alpha,\beta,\gamma,\delta$ is
said to be {\em $\delta$-improper} provided  that adjacent edges having the same colours (if any) are coloured with $\delta$. It is clear that a
proper edge colouring (and hence a $\delta$-minimum edge-colouring) of $G$ is a particular
$\delta$-improper edge colouring. For a proper or $\delta$-improper
edge colouring $\phi$ of $G$, it will be convenient to denote
$E_{\phi}(x)$  ($x \in \{\alpha,\beta,\gamma,\delta\})$ the set of
edges coloured with $x$ by $\phi$. For $x,y \in
\{\alpha,\beta,\gamma,\delta\}$,$x\neq y$, $\phi(x,y)$ is the partial subgraph
of $G$ spanned by these two colours, that is $E_{\phi}(x) \cup
E_{\phi}(y)$ (this subgraph being a union of paths and even cycles
where the colours $x$ and $y$ alternate). Since any two \dms of $G$
have the same number of edges coloured $\delta$ we shall denote by
$s(G)$ this number (the {\em colour number} as defined by Steffen in
\cite{Ste}).

 As usual, for any undirected graph  $G$, we denote by $V(G)$ the set of its vertices
and by $E(G)$ the set of its edges and we suppose that $|V(G)|=n$
and $|E(G)|=m$. Moreover, $V_i(G)$ denotes the set of vertices of $G$ of degree $i$, and when no confusion is possible we shall write $V_i$ instead of $V_i(G)$. A {\em strong matching} $C$ in a graph $G$ is a
matching $C$ such that there is no edge of $E(G)$ connecting any two
edges of $C$, or, equivalently, such that $C$ is the edge-set of the
subgraph of $G$ induced on the vertex-set $V(C)$.

\section{On \dm}\label{section:tools}
The graph $G$ considered in this section will have maximum degree $3$.

\begin{Lem} \label{Lemma:ImproperDelta} Let $\phi$ be a
$\delta-$improper colouring of $G$ then there exists a proper
colouring of $G$ $\phi'$ such that $E_{\phi'}(\delta) \subseteq
E_{\phi}(\delta)$
\end{Lem}
\begin{Prf}
Let $\phi$ be a $\delta$-improper edge colouring of $G$. If $\phi$
is a proper colouring, we are done. Hence, assume that $uv$ and $uw$
are coloured $\delta$. If $d(u)=2$ we can change the colour of $uv$
to $\alpha, \beta$ or $\gamma$ since $v$ is incident to at most two
colours in this set.

If $d(u)=3$ assume that the third edge $uz$ incident to $u$ is also
coloured $\delta$, then we can change the colour of $uv$ for the
same reason as above.

If $uz$ is coloured with $\alpha, \beta$ or $\gamma$, then $v$ and
$w$ are incident to the two remaining colours of the set $\{\alpha,\beta,\gamma\}$ otherwise one of the edges $uv$, $uw$ can
be recoloured with the missing colour. W.l.o.g., consider that $uz$
is coloured $\alpha$ then $v$ and $w$ are incident to $\beta$ and
$\gamma$. Since $u$ has degree $1$ in $\phi(\alpha,\beta)$ let $P$ be the path of $\phi(\alpha,\beta)$ which ends on $u$.   We can assume that $v$ or $w$ (say $v$)
is not the other end vertex of $P$. Exchanging $\alpha$ and $\beta$
along $P$ does not change the colours incident to $v$. But now $uz$
is coloured $\alpha$  and we can change the colour of $uv$ with
$\beta$.

In each case, we get hence a new $\delta$-improper edge colouring
$\phi_1$ with $E_{\phi_1}(\delta) \subsetneq E_{\phi}(\delta)$.
Repeating this process leads us to construct a proper edge colouring
of $G$ with $E_{\phi'}(\delta) \subseteq E_{\phi}(\delta)$ as
claimed.
\end{Prf}

\begin{Prop}\label{Proposition:Steffen1}
Let  $v_1,v_2, \ldots ,v_k \in V(G)$ be such that $G-\{v_1,v_2, \ldots,v_k\}$ is $3$-edge colourable. Then $s(G) \leq k$.
\end{Prop}
\begin{Prf}
Let us consider a $3$-edge colouring of $G-\{v_1,v_2, \ldots v_k\}$
with $\alpha, \beta$ and $\gamma$ and let us colour the edges
incident to $v_1,v_2, \ldots ,v_k$ with $\delta$. We get a
$\delta$-improper edge colouring $\phi$ of $G$. Lemma
\ref{Lemma:ImproperDelta} gives a proper colouring of $G$ $\phi'$
such that $E_{\phi'}(\delta) \subseteq E_{\phi}(\delta)$. Hence
$\phi'$ has at most $k$ edges coloured with $\delta$ and $s(G) \leq
k$.
\end{Prf}

Proposition \ref{Proposition:Steffen1} above has been obtained by
Steffen \cite{Ste} for cubic graphs.

\begin{Lem} \label{Lemma:ImproperDeltaSize} Let $\phi$ be a
$\delta-$improper colouring of $G$ then $|E_{\phi}(\delta)| \geq
s(G)$.
\end{Lem}
\begin{Prf}
Applying Lemma \ref{Lemma:ImproperDelta}, let $\phi'$ be a proper
edge colouring of $G$ such that $E_{\phi'}(\delta) \subseteq
E_{\phi}(\delta)$. We clearly have $|E_{\phi}(\delta)| \geq
|E_{\phi'}(\delta)| \geq s(G)$.
\end{Prf}

\begin{Thm}\label{Thm:Summarize}\cite{FouVan10a}
Let $G$ be a graph of maximum degree $3$ and $\phi$ be a $\delta$-minimum colouring of $G$.Then the following hold.\\
\begin{enumerate}
\item$E_{\phi}(\delta)=A_{\phi}\cup B_{\phi}\cup C_{\phi}$ where an edge $e$ in $A_{\phi}$ ($B_{\phi}$, $C_{\phi}$ respectively) belongs to a uniquely determined cycle
$C_{A_{\phi}}(e)$ ($C_{B_{\phi}}(e)$, $C_{C_{\phi}}(e)$ respectively) with precisely one edge coloured $\delta$ and the other edges being alternately coloured $\alpha$ 
and $\beta$ ($\beta$ and $\gamma$, $\alpha$ and $\gamma$ respectively). 
\item Each edge having exactly one vertex in common with some edge in $A_{\phi}$ ($B_{\phi}$, $C_{\phi}$ respectively) is coloured $\gamma$ ($\alpha$, $\beta$, respectively).
\item The multiset of colours of edges of $C_{A_{\phi}}(e)$ ($C_{B_{\phi}}(e)$, $C_{C_{\phi}}(e)$ respectively) can be permuted to obtain a (proper) $\delta$-minimum edge-colouring of G in which
the colour $\delta$ is moved from $e$ to an arbitrarily prescribed edge.
\item No two consecutive vertices of $C_{A_{\phi}}(e)$ ($C_{B_{\phi}}(e)$, $C_{C_{\phi}}(e)$ respectively) have degree 2.
\item The cycles from $1$ that correspond to distinct edges of $E_{\phi}(\delta)$ are vertex-disjoint.
\item If the edges $e_1, e_2, e_3 \in E_{\phi}(\delta)$ all belong to $A_{\phi}$ ($B_{\phi}$, $C_{\phi}$ respectively), then the set $\{e_1 , e_2 , e_3 \}$ 
induces in $G$ a subgraph with at most $4$ edges.
\end{enumerate}
\end{Thm}

\begin{Lem} \cite{FouPhD}\label{Lemma:OddCycleAssociated} 
Let $\phi$ be a $\delta$-minimum edge-colouring of $G$. For any edge
$e = uv \in E_{\phi}(\delta)$ with $d(v) \leq d(u)$ there is a colour $x  \in \{\alpha, \beta, \gamma\}$ present at
$v$ and a colour $y \in\{\alpha,\beta,\gamma\}-\{x\}$ present at $u$ such that one of connected
components of $\phi(x, y)$ is a path of even length joining the two ends of $e$.
Moreover, if $d(v)$ = 2, then both colours of $\{\alpha,\beta,\gamma\}-\{x\}$ satisfy the above
assertion.
\end{Lem}

An edge of $E_{\phi}(\delta)$ is in $A_{\phi}$ when its ends
can be connected by a path of $\phi(\alpha,\beta)$, $B_{\phi}$ by a
path of $\phi(\beta,\gamma)$ and $C_{\phi}$   by a path of
$\phi(\alpha,\gamma)$.
From Lemma  \ref{Lemma:OddCycleAssociated} it is clear that if $d(u) = 3$ and $d(v) = 2$ for an edge
$e = uv \in E_{\phi}(\delta)$, the $A_{\phi}$ , $B_{\phi}$ and $C_{\phi}$ are not pairwise disjoint; indeed, if the
colour $\gamma$ is present at the vertex $v$, then $e \in  A_{\phi}\cap B_{\phi}$.

When $e \in A_{\phi}$  we can associate to $e$ the odd cycle
$C_{A_{\phi}}(e)$  obtained by considering the
path of $\phi(\alpha,\beta)$  together with $e$. We define in the
same way $C_{B_{\phi}}(e)$ and $C_{C_{\phi}}(e)$ when $e$ is in
$B_{\phi}$ or $C_{\phi}$.

For each edge $e \in E_{\phi}(\delta)$ (where $\phi$ is a \dm of
$G$) we can associate one or two odd cycles following the fact
that $e$ is in one or two sets among $A_{\phi}$, $B_{\phi}$ or
$C_{\phi}$. Let ${\mathcal C}$ be the set of odd cycles associated
to edges in $E_{\phi}(\delta)$.

By Theorem \ref{Thm:Summarize} any two cycles in $\mathcal
C$ corresponding to edges in distinct sets $A_{\phi}$, $B_{\phi}$ or
$C_{\phi}$ are at distance at least $2$. Assume that
$C_1=C_{A_{\phi}}(e_1)$ and $C_2=C_{A_{\phi}}(e_2)$ for some edges
$e_1$ and $e_2$ in $A_{\phi}$. Can we say something about the
subgraph of $G$ whose vertex set is $V(C_1) \cup V(C_2)$~? In
general, we have no answer to this problem. However, when $G$ is
cubic and any vertex of $G$ lies on some cycle of $\mathcal C$ (we
shall say that $\mathcal C$ is {\em spanning}), we have a property
which will be useful later. Let us remark first that whenever
$\mathcal C$ is spanning, we can consider that $G$ is edge-coloured
in such a way that the edges of the cycles of $\mathcal C$ are
alternatively coloured with $\alpha$ and $\beta$ (except one edge
coloured $\delta$) and the remaining perfect matching is coloured
with $\gamma$. For this \dm of $G$ we have $B_{\phi} = \emptyset$ as
well as $C_{\phi}=\emptyset$.

\begin{Lem} \label{Lemma:DisjointOddCyclesInSameSet}
Assume that $G$ is cubic and $\mathcal C$ is spanning. Let $e_1,e_2
\in A_{\phi}$  and let $C_1,C_2 \in \mathcal C$ such that
$C_1=C_{A_{\phi}}(e_1)$ and $C_2=C_{A_{\phi}}(e_2)$.  Then at least
one of the following is true:

\begin{description}
\item[(i)] $C_1$ and $C_2$ are at distance at least $2$.
\item[(ii)] $C_1$ and $C_2$ are joined by at least $3$ edges.
\item[(iii)] $C_1$ and $C_2$ have at least two chords each.
\end{description}
\end{Lem}
\begin{Prf}
Since $e_1,e_2\in A_{\phi}$ and $\mathcal C$ is spanning we have $B_{\phi}=C_{\phi}=\emptyset$.
Let $C_1=v_0v_1 \ldots v_{2k_1}$ and $C_2=w_0w_1 \ldots w_{2k_2}$.
Assume that $C_1$ and $C_2$ are joined by the edge $v_0w_0$. By Theorem \ref{Thm:Summarize}, up to a re-colouring of the edges in $C_1$ and $C_2$, we can consider a \dm $\phi$
such that $\phi(v_0v_1)=\phi(w_0w_1)=\delta$,
$\phi(v_1v_2)=\phi(w_1w_2)=\beta$ and
$\phi(v_0v_{2k_1})=\phi(w_0w_{2k_2})=\alpha$. Moreover each edge of
$G$ (in particular $v_0w_0$) incident with these cycles is coloured
$\gamma$. We can change the colour of $v_0w_0$ in $\beta$. We obtain
thus a new \dm $\phi'$. Performing that exchange of colours on
$v_0w_0$ transforms the edges coloured $\delta$ $v_0v_1$ and
$w_0w_1$ in two edges of $C_{\phi'}$ lying on odd cycles $C'_1$ and
$C'_2$ respectively. We get hence a new set $\mathcal C'=\mathcal C
-\{C_1,C_2\}\cup\{C'_1,C'_2\}$ of odd cycles associated to
$\delta-$coloured edges in $\phi'$.

From Theorem \ref{Thm:Summarize}  $C'_1$ ($C'_2$
respectively) is at distance at least $2$ from any cycle in
$\mathcal C -\{C_1,C_2\}$. Hence $V(C'_1) \cup V(C'_2) \subseteq
V(C_1) \cup V(C_2)$. It is an easy task to check now that (ii) or
(iii) above must be verified.
\end{Prf}

\begin{Lem}\cite{FouPhD}\label{Lemma:OneVertexInNeighborhood}
Let $e_1=uv_1$ be an edge of $E_{\phi}(\delta)$ such that $v_1$ has
degree $2$ in $G$. Then $v_1$ is the only vertex in $N(u)$ of degree
 $2$ and for any other edge $e_2 \in E_{\phi}(\delta)$,  $\{e_1,e_2\}$ induces a $2K_2$.
\end{Lem}

\section{Applications}\label{section:Apps}
\subsection{On a result by Payan}
In \cite{Pay} Payan showed that it is always possible to edge-colour
a graph of maximum degree $3$ with three maximal matchings (with respect to the
inclusion) and introduced henceforth a notion of {\em strong-edge
colouring} where a strong edge-colouring means that one colour is a
strong matching while the remaining colours are usual matchings.
Payan conjectured that any $d-$regular graph has $d$ pairwise
disjoint maximal matchings and showed that this conjecture holds
true for graphs with maximum degree $3$.

The following result has been obtained first by Payan \cite{Pay}, but his
technique does not exhibit explicitly the odd cycles associated to
the edges of the strong matching and their properties.

\begin{Thm} \label{Theorem:Payan} Let $G$ be a graph with maximum degree at most $3$.
Then $G$ has a \dm $\phi$ where $E_{\phi}(\delta)$ is a strong
matching and, moreover, any edge in $E_{\phi}(\delta)$ has its two
ends of degree $3$ in $G$.
\end{Thm}
\begin{Prf}
Let $\phi$ be a \dm of $G$. From Theorem \ref{Thm:Summarize}, any two
edges of $E_{\phi}(\delta)$ belonging to distinct sets from among
$A_{\phi},B_{\phi}$ and $C_{\phi}$  are at least at distance $2$ and thus induce a strong matching. Hence,
we have to find a \dm where each set $A_{\phi}, B_{\phi}$ or
$C_{\phi}$ induces  a strong matching (with the supplementary
property that the end vertices of these edges have degree $3$). That
means that we can work on each set $A_{\phi}, B_{\phi}$ and
$C_{\phi}$ independently. Without loss of generality, we only consider $A_{\phi}$
here.

Assume that $A_{\phi}=\{e_1,e_2, \ldots e_k\}$ and
$A_{\phi}'=\{e_1,\ldots e_i\}$ ($1 \leq i \leq k-1$) is a strong
matching and each edge of $A_{\phi}'$ has its two ends with degree
$3$ in $G$. Consider the edge $e_{i+1}$ and let
$C=C_{e_{i+1}}(\phi)=u_0,u_1 \ldots u_{2p}$ be the odd cycle
associated to this edge (Theorem \ref{Thm:Summarize}).

Let us mark any vertex $v$ of degree $3$ on $C$ with a $+$ whenever
the edge of colour $\gamma$ incident to this vertex has its other
end which is a vertex incident to an edge of $A_{\phi}'$ and let us
mark $v$ with $-$ otherwise. By Theorem \ref{Thm:Summarize}, no consecutive vertices on $C$ have degree $2$, that means that a
vertex of degree $2$ on $C$ has its two neighbours of degree $3$ and
by Lemma \ref{Lemma:OneVertexInNeighborhood} these two vertices are marked with a
$-$. By Theorem \ref{Thm:Summarize} we cannot have two
consecutive vertices marked with a $+$, otherwise we would have three edges of $E_{\phi}(\delta)$ inducing a subgraph with more than $4$ edges, a contradiction. Hence, $C$ must have two
consecutive vertices of degree $3$ marked with $-$ whatever is the
number of vertices of degree $2$ on $C$.

Let $u_j$ and $u_{j+1}$ be two vertices of $C$ of degree $3$ marked
with $-$ ($j$ being taken modulo $2p+1$). We can transform the edge
colouring $\phi$ by exchanging colours on $C$ uniquely, in such a
way that the edge of colour $\delta$ of this cycle is $u_ju_{j+1}$.
In the resulting edge colouring $\phi_1$ we have
$A_{\phi_1}=A_{\phi}-\{e_{i+1}\}\cup\{u_ju_{j+1}\}$ and
$A_{\phi_1}'=A_{\phi}'\cup\{u_ju_{j+1}\}$ is a strong matching where each
edge has its two ends of degree $3$. Repeating this process we are
left with a new  $\delta$-minimum colouring $\phi'$ where $A_{\phi'}$ is a strong matching.
\end{Prf}

\begin{Cor} \label{Corollary:Subgraph3Edgecolourable}
Let $G$ be a  graph  with maximum degree $3$ then there are $s(G)$
vertices of degree $3$ pairwise non-adjacent $v_1 \ldots v_{s(G)}$
such that $G -\{v_1 \ldots v_{s(G)}\}$ is $3$-colourable.
\end{Cor}
\begin{Prf}
Pick a vertex on each edge coloured $\delta$ in a $\delta$-minimum
colouring $\phi$ of $G$ where $E_{\phi}(\delta)$ is a strong
matching (Theorem \ref{Theorem:Payan}). We get a subset $S$ of
vertices satisfying our corollary.
\end{Prf}

Steffen \cite{Ste} obtained Corollary
\ref{Corollary:Subgraph3Edgecolourable} for bridgeless cubic graphs.

\subsection{Parsimonious edge colouring}
Let $\chi'(G)$ be the classical chromatic index of $G$. For
convenience let
\begin{equation*}
c(G)=max\{|E(H)|: H \subseteq G \\\ ,\chi'(H)=3\}
\end{equation*}
\begin{equation*}
\gamma(G)=\frac{c(G)}{|E(G)|}
\end{equation*}
Staton \cite{Sta} (and independently Locke \cite{Loc}) showed that
whenever $G$ is a cubic graph distinct from $K_4$ then $G$ contains
a bipartite subgraph (and hence a $3$-edge colourable graph, by
K\"{o}nig's theorem \cite{Kon}) with at least $\frac{7}{9}$ of the
edges of $G$. Bondy and Locke \cite{BonLock} obtained $\frac{4}{5}$
when considering graphs with maximum degree at most $3$.

In \cite{AlbHaa} Albertson and Haas showed that whenever $G$ is a
cubic graph, we have $\gamma(G) \geq \frac{13}{15}$ while for graphs
with maximum degree $3$ they obtained $\gamma(G) \geq
\frac{26}{31}$. Our purpose here is to show that $\frac{13}{15}$  is
a lower bound for $\gamma(G)$ when $G$ has maximum degree $3$, with
the exception of the graph $G_5$ depicted in  Figure
\ref{Figure:Gfive} below.
\begin{figure}[htb]
\begin{center}
\setlength{\unitlength}{0.00034996in}
\begingroup\makeatletter\ifx\SetFigFont\undefined
\def\x#1#2#3#4#5#6#7\relax{\def\x{#1#2#3#4#5#6}}%
\expandafter\x\fmtname xxxxxx\relax \def\y{splain}%
\ifx\x\y   
\gdef\SetFigFont#1#2#3{%
  \ifnum #1<17\tiny\else \ifnum #1<20\small\else
  \ifnum #1<24\normalsize\else \ifnum #1<29\large\else
  \ifnum #1<34\Large\else \ifnum #1<41\LARGE\else
     \huge\fi\fi\fi\fi\fi\fi
  \csname #3\endcsname}%
\else
\gdef\SetFigFont#1#2#3{\begingroup
  \count@#1\relax \ifnum 25<\count@\count@25\fi
  \def\x{\endgroup\@setsize\SetFigFont{#2pt}}%
  \expandafter\x
    \csname \romannumeral\the\count@ pt\expandafter\endcsname
    \csname @\romannumeral\the\count@ pt\endcsname
  \csname #3\endcsname}%
\fi
\fi\endgroup
{\renewcommand{\dashlinestretch}{30}
\begin{picture}(4066,3856)(0,-10)
\put(2933,3608){\blacken\ellipse{450}{450}}
\put(2933,3608){\ellipse{450}{450}}
\put(3833,1808){\blacken\ellipse{450}{450}}
\put(3833,1808){\ellipse{450}{450}}
\put(233,1808){\blacken\ellipse{450}{450}}
\put(233,1808){\ellipse{450}{450}}
\put(2033,233){\blacken\ellipse{450}{450}}
\put(2033,233){\ellipse{450}{450}}
\put(1133,3608){\blacken\ellipse{450}{450}}
\put(1133,3608){\ellipse{450}{450}}
\path(233,1808)(2933,3608)
\path(3833,1808)(1133,3608)(2933,3608)
	(3833,1808)(2033,233)(233,1808)(1133,3608)
\end{picture}
}
\end{center}
\caption{$G_5$} \label{Figure:Gfive}
\end{figure}

\begin{Lem} \label{Lemma:bG} Let $G$ be a graph with maximum degree $3$ then $\gamma(G)= 1-\frac{s(G)}{m}$.
\end{Lem}
\begin{Prf}
Let $\phi$ be a \dm of $G$. The restriction of $\phi$ to $E(G)-E_{\phi}(\delta)$ is a proper 3-edge-colouring, hence $c(G)\geq m-s(G)$ and $\gamma(G)\geq 1-\frac{s(G)}{m}$.

If $H$ is a subgraph of $G$ with $\chi(H)=3$, consider a proper 3-edge-colouring $\phi : E(H) \rightarrow \{\alpha,\beta,\gamma\}$ and let $\psi : E(G) \rightarrow \{\alpha,\beta,\gamma,\delta\}$ be the continuation of $\phi$ with $\psi(e)=\delta$ for $e\in E(G)-E(H)$. By Lemma \ref{Lemma:ImproperDelta} there is a proper edge-colouring  $\psi'$ of $G$ with $E_{\psi'}(\delta)\subseteq E_{\psi}(\delta)$  so that $|E(H)|=|E(G)-E_{\psi}(\delta)|\leq |E(G)-E_{\psi'}(\delta)|\leq m-s(G)$,
$c(G)\leq m-s(G)$ and $\gamma(G)\leq 1-\frac{s(G)}{m}$.
\end{Prf}

In \cite{Riz09}, Rizzi shows that for triangle-free graphs of maximum degree $3$, $\gamma(G)\geq 1-\frac{2}{3g_{o}(G)}$  (where the {\em odd girth} of a graph $G$,
denoted by $g_{o}(G)$, is the minimum length of an odd cycle in $G$).

\begin{Thm}\label{Thm:RizziPlus}
Let $G$ be a graph with maximum degree $3$ then $\gamma(G)\geq 1-\frac{2}{3g_{o}(G)}$.
\end{Thm}
\begin{Prf}
Let $\phi$ be a \dm of $G$ and $E_{\phi}(\delta)=\{e_1,e_2\ldots,e_{s(G)}\}$. ${\mathcal C}$ being the set of odd cycles associated to edges in $E_{\phi}(\delta)$, 
we write $\mathcal C=\{C_1,C_2\ldots,C_{s(G)}\}$ and suppose that for $i=1,2\ldots,s(G)$, $e_i$ is an edge of $C_i$. 
We know by Theorem \ref{Thm:Summarize} that the cycles of $\mathcal C$ are vertex-disjoint.

Let us write $\mathcal C=\mathcal C_2 \cup \mathcal C_3$, where $\mathcal C_2$ denotes the set of odd cycles of $\mathcal C$ which have a vertex of degree $2$, 
while $\mathcal C_3$ is for the set of cycles in $\mathcal C$ whose all vertices have degree $3$. 
Let $k=|\mathcal C_2|$, obviously we have $0\leq k\leq s(G)$ and $\mathcal C_2\cap\mathcal C_3=\emptyset$.

If $C_i\in\mathcal C_2$ we suppose without loss of generality that $C_i\in A_{\phi}$ and we have $|C_i|\geq g_o(G)$. Moreover,  since any edge in $C_i$ can be coloured 
$\delta$  (Theorem \ref{Thm:Summarize}), we may assume that $e_i$ has a vertex of degree $2$. We can associate to $e_i$ another odd cycle say $C^{'}_i\in B_{\phi}$
 (Lemma \ref{Lemma:OddCycleAssociated}) whose edges distinct from $e_i$ form  an even path of $\phi(\alpha,\gamma)$ using at least $\frac{g_{o}(G)-1}{2}$ edges, 
coloured $\gamma$, which are not edges of $C_i$. 

When $|C_i|>g_o(G)$ or $|C'_i|>g_o(G)$ there are either at least $g_o(G)+2$ edges in $C_i$ or at least $\frac{g_{o}(G)-1}{2}+1$ edges  coloured $\gamma$ in $C'_i$.
If $|C_i|=|C'_i|=g_o(G)$ there is at least one edge coloured $\alpha$ in $C'_i$ that is not an edge of $C_i$, otherwise all the edges coloured $\gamma$ of $C'_i$ would 
be chords of $C_i$, a contradiction since a such chord would form with vertices of $C_i$ an odd cycle of length smaller than $g_o(G)$.

Hence, $C_i\cup C^{'}_i$ contains at least $g_o(G)+\frac{g_{o}(G)-1}{2}+1> \frac{3}{2}g_o(G)$ edges. 

Consequently there are at least $\frac{3}{2}\times k\times g_{o}(G)$ edges in $\displaystyle\bigcup_{C_i\in\mathcal C_2}(C_i\cup C^{'}_i)$.

When $C_i\in\mathcal C_3$, $C_i$ contains at least $g_{o}(G)$ edges, moreover, each vertex of $C_i$ being of degree $3$, 
there are at least $\frac{s(G)-k}{2}\times g_{o}(G)$ additionnal edges which are incident to a vertex of $\displaystyle\bigcup_{C_i\in\mathcal C_3} C_{i}$.

Since $C_i\cap C_j=\emptyset$ and $C^{'}_i\cap C_j=\emptyset$ ($1\leq i,j\leq s(G)$, $i\neq j$), we have
$$m\geq \frac{3}{2}g_{o}(G)\times k +(s(G)-k)\times g_{o}(G)+\frac{s(G)-k}{2}\times g_{o}(G)=\frac{3}{2}\times s(G)\times g_{o}(G).$$

Consequently $\gamma(G)= 1-\frac{s(G)}{m}\geq 1-\frac{2}{3g_{o}(G)}$.
\end{Prf}

As a matter of fact, $\gamma(G)> 1-\frac{2}{3g_{o}(G)}$ when the graph $G$ contains vertices of degree $2$. 
In a further work (see \cite{FouVan11a}) we refine the bound and prove that $\gamma(G)\geq 1-\frac{2}{3g_{o}(G)+2}$ when $G$ is a graph of maximal degree $3$ distinct from the Petersen graph.
\begin{Lem} \label{Lemma:NoPendantVertex} \cite{AlbHaa} Let $G$ be a graph with maximum degree $3$.
Assume that $v \in V(G)$ is such that $d(v)=1$ then $\gamma(G) > \gamma(G-v)$.
\end{Lem}

A {\em triangle} $T=\{a,b,c\}$ is said to be {\em reducible}
whenever its neighbours are distinct. When $T$ is a reducible
triangle in $G$ ($G$ having maximum degree $3$)  we can obtain a new
graph $G'$ with maximum degree $3$ by shrinking this triangle into a
single vertex and joining this new vertex to the neighbours of $T$ in
$G$.

\begin{Lem} \label{Lemma:NoReducibleTriangle} \cite{AlbHaa} Let $G$ be a graph with maximum degree $3$.
Assume that $T=\{a,b,c\}$ is a reducible triangle and let $G'$ be
the graph obtained by reduction of this triangle. Then
 $\gamma(G) > \gamma(G')$.
\end{Lem}

\begin{Thm} \label{Theorem:Gamma}
Let $G$ be a  graph  with maximum degree $3$. If $G \neq G_5$ then $\gamma(G) \geq 1-\frac{\frac{2}{15}}{1+\frac{2}{3}\frac{|V_2|}{|V_3|} }$.
\end{Thm}
\begin{Prf}  From Lemma \ref{Lemma:NoPendantVertex} and Lemma \ref{Lemma:NoReducibleTriangle} we can consider that $G$ has only vertices of degree
$2$ or $3$ and that $G$ contains no reducible triangle. 

Assume that we can associate a set $P_e$ of at least $5$ distinct
vertices of  $V_3$ for each edge $e \in E_{\phi}(\delta)$ in a \dm
$\phi$ of $G$. Assume moreover that
\begin{equation} \label{Equation:PrivateV3}
\forall e,e' \in E_{\phi}(\delta) \ \ P_e \cap P_{e'} = \emptyset
\end{equation}
Then
\begin{equation*}
\gamma(G)= 1-\frac{s(G)}{m}=1-\frac{s(G)}{\frac{3}{2}|V_3|+|V_2|} \geq 1-\frac{\frac{|V_3|}{5}}{\frac{3}{2}|V_3|+|V_2|}
\end{equation*}
Hence
\begin{equation*}
\gamma(G) \geq 1-\frac{\frac{2}{15}}{1+\frac{2}{3}\frac{|V_2|}{|V_3|} }
\end{equation*}

It remains to see how to construct the sets  $P_e$ satisfying
Property (\ref{Equation:PrivateV3}). Let ${\mathcal C}$ be the set
of odd cycles associated to edges in $E_{\phi}(\delta)$. Let $e \in E_{\phi}(\delta)$, assume
that $e$ is contained in a cycle $C \in \mathcal C$ of length $3$.
By Theorem \ref{Thm:Summarize}, the  edges incident to that
triangle have the same colour in $\{\alpha, \beta,\gamma\}$. This
triangle is hence reducible, impossible. We can thus consider that
each cycle of $\mathcal C$ has length at least $5$. By Lemma
\ref{Lemma:OneVertexInNeighborhood}, we know that whenever such a
cycle contains vertices of $V_2$, their distance on this  cycle is
at least $3$. Which means that every cycle $C \in \mathcal C$
contains at least $5$ vertices in $V_3$ as soon as $C$ has length at
least $7$ or $C$ has length $5$ but does not contain a vertex of
$V_2$. For each edge $e \in E_{\phi}(\delta)$ contained in such a
cycle we associate $P_e$ as any set of $5$ vertices of $V_3$
contained in the cycle.

There may exist edges in $E_{\phi}(\delta)$ contained in a $5$-cycle
of $\mathcal C$ having exactly one vertex in $V_2$. Let
$C=a_1a_2a_3a_4a_5$ be such a  cycle and assume that $a_1 \in V_2$.
By Lemma \ref{Lemma:OneVertexInNeighborhood}, $a_1$ is the only vertex of degree $2$
and by exchanging colours along this cycle, we
can suppose that $a_1a_2 \in E_{\phi}(\delta)$. Since $a_1 \in V_2$,
$e=a_1a_2$ is contained in a second cycle $C'$ of $\mathcal C$ (see
Remark \ref{Remark:NoUnicityFondamentalOddCycle}). If $C'$ contains
a vertex $ x \in V_3$ distinct from $a_2,a_3,a_4$ and $a_5$ then we
set $P_e=\{a_2,a_3,a_4,a_5,x\}$. Otherwise $C'=a_1a_2a_4a_3a_5$ and
$G$ is isomorphic to $G_5$, impossible.

The sets $\{P_e |\;  e \in E_{\phi}(\delta)\}$ are pairwise disjoint
since any two cycles of $\mathcal C$ associated to distinct edges in
$E_{\phi}(\delta)$ are disjoint. Hence Property
\ref{Equation:PrivateV3} holds and the proof is complete.
\end{Prf}

Albertson and Haas \cite{AlbHaa} proved that $\gamma(G) \geq
\frac{26}{31}$ when $G$ is a graph with maximum degree $3$ and Rizzi \cite{Riz09} obtained $\gamma(G)\geq \frac{6}{7}$. 
From Theorem \ref{Theorem:Gamma} we get immediately for all graphs $G\neq G_5$ $\gamma(G)\geq \frac{13}{15}$, a better bound. 
Let us remark that we get also $\gamma(G) \geq \frac{13}{15}$ by Theorem \ref{Thm:RizziPlus} as soon as $g_{o}(G) \geq 5$.

\begin{Lem}\label{Lemmma:DeuxFacteurDeC5QuiAtteintLeMinimum}
Let $G$ be a cubic graph which can be factored into $s(G)$ cycles of
length $5$ and has no reducible triangle. Then every $2$-factor of
$G$ contains $s(G)$ cycles of length $5$.
\end{Lem}
\begin{Prf}
Since $G$ has no reducible triangle, all cycles in a $2$-factor have
length at least $4$. Let $\mathcal{C}$ be any $2$-factor of $G$.
Let us denote $n_4$ the number of cycles of length $4$, $n_5$ the number of cycles of length $5$ and $n_{6+}$
the number of cycles on at least $6$ vertices in $\mathcal{C}$.
We have $5n_5+6n_{6+}\leq 5s(G)-4n_4$.

If $n_4+n_ {6_+}=0$, then $n_5=s(G)$. If $n_4+n_ {6_+}> 0$, then the number of odd cycles in $\mathcal{C}$ 
is at most $n_5+n_{6_+}\leq \frac{5s(G)-4n_4-n_{6_+}}{5}=\frac{5s(G)-(n_4+n_{6_+})-3n_4}{5}<s(G)$.
A contradiction since a $2$-factor of $G$ contains at least $s(G)$ odd cycles.
\end{Prf}

\begin{Cor} \label{Corollary:CubicExtremal}
Let $G$ be a  graph with maximum degree $3$ such that $\gamma(G) =
\frac{13}{15}$. Then $G$ is a cubic graph which can be factored into
$s(G)$  cycles of length $5$. Moreover every $2$-factor of $G$ has
this property.
\end{Cor}
\begin{Prf}The optimum for $\gamma(G)$ in Theorem \ref{Theorem:Gamma} is obtained whenever
$s(G) = \frac{|V_3|}{5}$ and $|V_2|=0$. That is, $G$ is a cubic
graph admitting a $2$-factor of $s(G)$  cycles of length $5$.
Moreover by Lemma \ref{Lemma:NoReducibleTriangle} $G$ has no
reducible triangle, the result comes from Lemma
\ref{Lemmma:DeuxFacteurDeC5QuiAtteintLeMinimum}.
\end{Prf}

As pointed out by Albertson and Haas \cite{AlbHaa}, the Petersen
graph with $\gamma(G) = \frac{13}{15}$ supplies an extremal example
for cubic graphs.  Steffen \cite{Ste04} proved that the only cubic
bridgeless graph with $\gamma(G) = \frac{13}{15}$ is the Petersen
graph. In fact, we  can extend this result to graphs with maximum
degree $3$ where bridges are allowed (excluding the graph $G_5$).
Let $P'$ be the cubic graph on $10$ vertices obtained from two
copies of $G_5$ (Figure \ref{Figure:Gfive}) by
joining by an edge the two vertices of degree $2$.
\begin{Thm} \label{Thm:OnlyPetersen}
Let $G$ be a connected graph with maximum degree $3$ such that
$\gamma(G)=\frac{13}{15}$. Then $G$ is isomorphic to the Petersen
graph or to $P'$.
\end{Thm}

\begin{Prf}Let $G$ be a graph with maximum degree $3$ such that $\gamma(G)=\frac{13}{15}$.

From Corollary \ref{Corollary:CubicExtremal}, we can consider that
$G$ is cubic and $G$ has a $2$-factor of  cycles of length $5$. Let
$\mathcal  C =\{C_1 \ldots C_{s(G)}\}$ be such a $2$-factor (
$\mathcal C$ is spanning). Let $\phi$ be a \dm of $G$ induced by
this $2-$factor.

Without loss of generality consider two cycles in $\mathcal{C}$,
namely $C_1$ and $C_2$, and let us  denote $C_1=v_1v_2v_3v_4v_5$
while $C_2=u_1u_2u_3u_4u_5$ and assume that $v_1u_1\in G$. From
Lemma \ref{Lemma:DisjointOddCyclesInSameSet}, $C_1$ and $C_2$ are
joined by at least $3$ edges or  each of them has two chords. If
$s(G)>2$ there is a cycle $C_3 \in \mathcal C$. Without loss of
generality, $G$ being connected, we can suppose that $C_3$ is joined
to $C_1$ by an edge. Applying one more time Lemma
\ref{Lemma:DisjointOddCyclesInSameSet}, $C_1$ and $C_3$ have two
chords or are joined by at least $3$ edges, contradiction with the
constraints imposed by $C_1$ and $C_2$. Hence $s(G)=2$ and $G$ has
$10$ vertices and no $4$-cycle, which leads to a graph isomorphic to
$P'$ or the Petersen graph as claimed.
\end{Prf}

We can construct  cubic graphs with chromatic index $4$ ({\em snarks} in the litterature) which are cyclically $4$- edge connected and having a $2$-factor of $C_5$'s.

Indeed, let $G$ be a cubic cyclically $4$-edge connected graph of order $n$ and $M$ be a perfect matching of $G$, $M=\{x_iy_i | i=1\ldots \frac{n}{2}\}$. 
Let $P_1\ldots P_{\frac{n}{2}}$ be $\frac{n}{2}$ copies of the Petersen graph. For each $P_i$ ($i=1\ldots \frac{n}{2}$) we consider two edges at distance 1 apart $e_i^1$ and  $e_i^2$. Let us observe that $P_i-\{e_i^1,e_i^2\}$ contains a $2$-factor of two $C_5$'s ($C_i^1$ and $C_i^2$).

\begin{figure}
 \begin{center}
  \setlength{\unitlength}{0.00034996in}
\begingroup\makeatletter\ifx\SetFigFont\undefined%
\gdef\SetFigFont#1#2#3#4#5{%
  \reset@font\fontsize{#1}{#2pt}%
  \fontfamily{#3}\fontseries{#4}\fontshape{#5}%
  \selectfont}%
\fi\endgroup%
{\renewcommand{\dashlinestretch}{30}
\begin{picture}(8754,11512)(0,-10)
\put(2172,4735){\blacken\ellipse{202}{202}}
\put(2172,4735){\ellipse{202}{202}}
\put(2172,4240){\blacken\ellipse{202}{202}}
\put(2172,4240){\ellipse{202}{202}}
\put(2172,1765){\blacken\ellipse{202}{202}}
\put(2172,1765){\ellipse{202}{202}}
\put(2172,1270){\blacken\ellipse{202}{202}}
\put(2172,1270){\ellipse{202}{202}}
\put(6537,4735){\blacken\ellipse{202}{202}}
\put(6537,4735){\ellipse{202}{202}}
\put(6492,1765){\blacken\ellipse{202}{202}}
\put(6492,1765){\ellipse{202}{202}}
\put(6537,4240){\blacken\ellipse{202}{202}}
\put(6537,4240){\ellipse{202}{202}}
\put(6492,1281){\blacken\ellipse{202}{202}}
\put(6492,1281){\ellipse{202}{202}}
\put(117,1150){\arc{210}{1.5708}{3.1416}}
\put(117,5080){\arc{210}{3.1416}{4.7124}}
\put(2787,5080){\arc{210}{4.7124}{6.2832}}
\put(2787,1150){\arc{210}{0}{1.5708}}
\path(12,1150)(12,5080)
\path(117,5185)(2787,5185)
\path(2892,5080)(2892,1150)
\path(2787,1045)(117,1045)
\put(5922,1150){\arc{210}{1.5708}{3.1416}}
\put(5922,5080){\arc{210}{3.1416}{4.7124}}
\put(8592,5080){\arc{210}{4.7124}{6.2832}}
\put(8592,1150){\arc{210}{0}{1.5708}}
\path(5817,1150)(5817,5080)
\path(5922,5185)(8592,5185)
\path(8697,5080)(8697,1150)
\path(8592,1045)(5922,1045)
\path(2172,4735)(6492,4735)
\path(2172,4240)(6537,4240)
\path(2172,1765)(6492,1765)
\path(2172,1270)(6492,1270)
\put(2217,11035){\blacken\ellipse{202}{202}}
\put(2217,11035){\ellipse{202}{202}}
\put(2217,10540){\blacken\ellipse{202}{202}}
\put(2217,10540){\ellipse{202}{202}}
\put(2217,8065){\blacken\ellipse{202}{202}}
\put(2217,8065){\ellipse{202}{202}}
\put(2217,7570){\blacken\ellipse{202}{202}}
\put(2217,7570){\ellipse{202}{202}}
\put(6582,11035){\blacken\ellipse{202}{202}}
\put(6582,11035){\ellipse{202}{202}}
\put(6582,10495){\blacken\ellipse{202}{202}}
\put(6582,10495){\ellipse{202}{202}}
\put(6537,8065){\blacken\ellipse{202}{202}}
\put(6537,8065){\ellipse{202}{202}}
\put(6537,7615){\blacken\ellipse{202}{202}}
\put(6537,7615){\ellipse{202}{202}}
\put(3432,10765){\blacken\ellipse{202}{202}}
\put(3432,10765){\ellipse{202}{202}}
\put(3387,7840){\blacken\ellipse{202}{202}}
\put(3387,7840){\ellipse{202}{202}}
\put(162,7450){\arc{210}{1.5708}{3.1416}}
\put(162,11380){\arc{210}{3.1416}{4.7124}}
\put(2832,11380){\arc{210}{4.7124}{6.2832}}
\put(2832,7450){\arc{210}{0}{1.5708}}
\path(57,7450)(57,11380)
\path(162,11485)(2832,11485)
\path(2937,11380)(2937,7450)
\path(2832,7345)(162,7345)
\put(5967,7450){\arc{210}{1.5708}{3.1416}}
\put(5967,11380){\arc{210}{3.1416}{4.7124}}
\put(8637,11380){\arc{210}{4.7124}{6.2832}}
\put(8637,7450){\arc{210}{0}{1.5708}}
\path(5862,7450)(5862,11380)
\path(5967,11485)(8637,11485)
\path(8742,11380)(8742,7450)
\path(8637,7345)(5967,7345)
\path(2217,11035)(3387,10765)
\path(2217,10540)(3432,10765)
\path(2217,8065)(3387,7840)
\path(2217,7570)(3387,7840)
\path(3432,10765)(3387,7840)
\path(6582,11035)(6581,11035)(6578,11036)
	(6573,11036)(6564,11037)(6553,11039)
	(6538,11041)(6519,11043)(6497,11045)
	(6471,11048)(6442,11052)(6411,11055)
	(6378,11058)(6343,11061)(6306,11065)
	(6269,11067)(6230,11070)(6191,11072)
	(6152,11074)(6112,11075)(6071,11076)
	(6030,11076)(5988,11075)(5945,11073)
	(5902,11070)(5858,11065)(5814,11060)
	(5769,11053)(5725,11045)(5682,11035)
	(5632,11021)(5586,11006)(5547,10991)
	(5513,10976)(5485,10963)(5462,10950)
	(5443,10939)(5428,10929)(5416,10920)
	(5407,10912)(5400,10904)(5393,10896)
	(5388,10889)(5382,10880)(5377,10871)
	(5371,10861)(5364,10850)(5357,10837)
	(5349,10822)(5341,10805)(5333,10786)
	(5327,10765)(5323,10743)(5322,10720)
	(5326,10698)(5334,10677)(5344,10659)
	(5354,10643)(5365,10629)(5375,10618)
	(5384,10608)(5392,10601)(5400,10594)
	(5406,10589)(5413,10585)(5420,10581)
	(5427,10578)(5435,10574)(5446,10570)
	(5459,10565)(5475,10559)(5495,10553)
	(5519,10545)(5550,10536)(5585,10526)
	(5627,10516)(5675,10505)(5727,10495)
	(5771,10488)(5817,10482)(5862,10477)
	(5908,10473)(5952,10470)(5996,10468)
	(6039,10467)(6081,10466)(6122,10466)
	(6162,10466)(6202,10467)(6242,10468)
	(6280,10470)(6318,10472)(6356,10474)
	(6392,10476)(6426,10478)(6459,10481)
	(6490,10483)(6518,10485)(6543,10488)
	(6565,10489)(6583,10491)(6598,10492)
	(6610,10493)(6618,10494)(6623,10495)
	(6626,10495)(6627,10495)
\path(6475,8124)(6474,8124)(6471,8125)
	(6466,8125)(6457,8126)(6446,8128)
	(6431,8130)(6412,8132)(6390,8134)
	(6364,8137)(6335,8141)(6304,8144)
	(6271,8147)(6236,8150)(6199,8154)
	(6162,8156)(6123,8159)(6084,8161)
	(6045,8163)(6005,8164)(5964,8165)
	(5923,8165)(5881,8164)(5838,8162)
	(5795,8159)(5751,8154)(5707,8149)
	(5662,8142)(5618,8134)(5575,8124)
	(5525,8110)(5479,8095)(5440,8080)
	(5406,8065)(5378,8052)(5355,8039)
	(5336,8028)(5321,8018)(5309,8009)
	(5300,8001)(5293,7993)(5286,7985)
	(5281,7978)(5275,7969)(5270,7960)
	(5264,7950)(5257,7939)(5250,7926)
	(5242,7911)(5234,7894)(5226,7875)
	(5220,7854)(5216,7832)(5215,7809)
	(5219,7787)(5227,7766)(5237,7748)
	(5247,7732)(5258,7718)(5268,7707)
	(5277,7697)(5285,7690)(5293,7683)
	(5299,7678)(5306,7674)(5313,7670)
	(5320,7667)(5328,7663)(5339,7659)
	(5352,7654)(5368,7648)(5388,7642)
	(5412,7634)(5443,7625)(5478,7615)
	(5520,7605)(5568,7594)(5620,7584)
	(5664,7577)(5710,7571)(5755,7566)
	(5801,7562)(5845,7559)(5889,7557)
	(5932,7556)(5974,7555)(6015,7555)
	(6055,7555)(6095,7556)(6135,7557)
	(6173,7559)(6211,7561)(6249,7563)
	(6285,7565)(6319,7567)(6352,7570)
	(6383,7572)(6411,7574)(6436,7577)
	(6458,7578)(6476,7580)(6491,7581)
	(6503,7582)(6511,7583)(6516,7584)
	(6519,7584)(6520,7584)
\put(1632,4690){\makebox(0,0)[lb]{\smash{{\SetFigFont{5}{6.0}{\rmdefault}{}{}$r$}}}}
\put(1632,4195){\makebox(0,0)[lb]{\smash{{\SetFigFont{5}{6.0}{\rmdefault}{}{}$s$}}}}
\put(6807,4690){\makebox(0,0)[lb]{\smash{{\SetFigFont{5}{6.0}{\rmdefault}{}{}$a$}}}}
\put(6807,4195){\makebox(0,0)[lb]{\smash{{\SetFigFont{5}{6.0}{\rmdefault}{}{}$b$}}}}
\put(1677,1720){\makebox(0,0)[lb]{\smash{{\SetFigFont{5}{6.0}{\rmdefault}{}{}$t$}}}}
\put(1632,1225){\makebox(0,0)[lb]{\smash{{\SetFigFont{5}{6.0}{\rmdefault}{}{}$u$}}}}
\put(6807,1720){\makebox(0,0)[lb]{\smash{{\SetFigFont{5}{6.0}{\rmdefault}{}{}$c$}}}}
\put(6852,1225){\makebox(0,0)[lb]{\smash{{\SetFigFont{5}{6.0}{\rmdefault}{}{}$d$}}}}
\put(1677,10990){\makebox(0,0)[lb]{\smash{{\SetFigFont{5}{6.0}{\rmdefault}{}{}$r$}}}}
\put(3702,10720){\makebox(0,0)[lb]{\smash{{\SetFigFont{5}{6.0}{\rmdefault}{}{}$x$}}}}
\put(3612,7750){\makebox(0,0)[lb]{\smash{{\SetFigFont{5}{6.0}{\rmdefault}{}{}$y$}}}}
\put(1677,10495){\makebox(0,0)[lb]{\smash{{\SetFigFont{5}{6.0}{\rmdefault}{}{}$s$}}}}
\put(1722,8020){\makebox(0,0)[lb]{\smash{{\SetFigFont{5}{6.0}{\rmdefault}{}{}$t$}}}}
\put(1722,7570){\makebox(0,0)[lb]{\smash{{\SetFigFont{5}{6.0}{\rmdefault}{}{}$u$}}}}
\put(6897,10990){\makebox(0,0)[lb]{\smash{{\SetFigFont{5}{6.0}{\rmdefault}{}{}$a$}}}}
\put(6852,10450){\makebox(0,0)[lb]{\smash{{\SetFigFont{5}{6.0}{\rmdefault}{}{}$b$}}}}
\put(6852,8020){\makebox(0,0)[lb]{\smash{{\SetFigFont{5}{6.0}{\rmdefault}{}{}$c$}}}}
\put(6807,7570){\makebox(0,0)[lb]{\smash{{\SetFigFont{5}{6.0}{\rmdefault}{}{}$d$}}}}
\put(4017,100){\makebox(0,0)[lb]{\smash{{\SetFigFont{8}{9.6}{\rmdefault}{}{}$G_1 . G_2$}}}}
\put(777,6670){\makebox(0,0)[lb]{\smash{{\SetFigFont{8}{9.6}{\rmdefault}{}{}$G_1$}}}}
\put(6582,6760){\makebox(0,0)[lb]{\smash{{\SetFigFont{8}{9.6}{\rmdefault}{}{}$G_2$}}}}
\end{picture}
}
  \caption{The dot product operation on graphs $G_1$, $G_2$.}\label{fig:DotProduct}
 \end{center}

\end{figure}
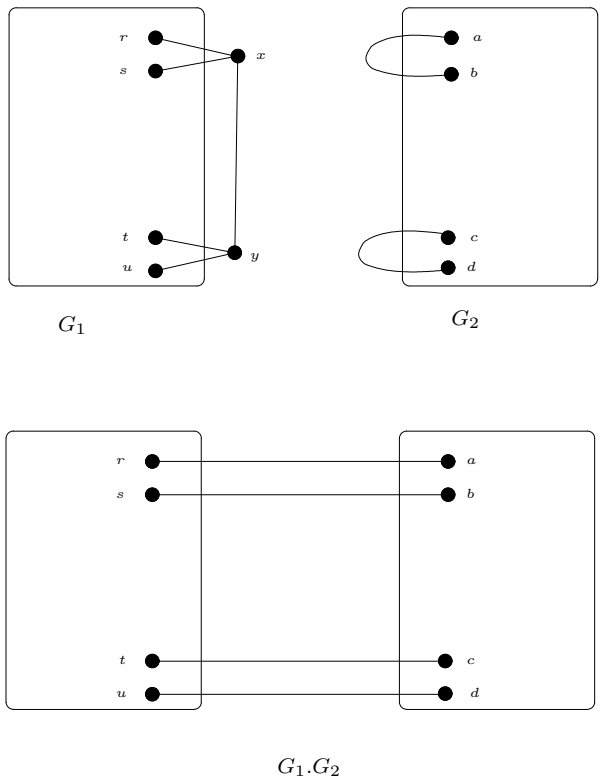

We construct then a new cyclically $4$-edge connected cubic graph $H$ with chromatic index $4$ by applying the well known operation dot-product 
(see Figure \ref{fig:DotProduct}, see Isaacs \cite{Isa75} for a description and for a formal definition) on $\{e_i^1,e_i^2\}$ 
and the edge $x_iy_i$ ($i=1\ldots \frac{n}{2}$). We remark that the vertices of $G$ vanish in the operation and the resulting graph $H$ has a $2$ factor of $C_5$, namely
$\{C_1^1,C_1^2,\ldots C_i^1,C_i^2,\ldots C_{\frac{n}{2}}^1, C_{\frac{n}{2}}^2\}$.

We do not know an example 
of a cyclically $5$-edge connected snark (except the Petersen graph)  with a $2$-factor of induced cycles of length $5$.

\begin{Problem} Is there any cyclically $5$-edge connected snark distinct from the Petersen graph with a $2$-factor of $C_5$'s~?
\end{Problem}

As a first step towards the resolution of this Problem we
propose the following Theorem. Recall that a {\em permutation graph} is a cubic graph having some $2$-factor with precisely $2$ odd cycles.

\begin{Thm}\label{thm:PasDe2FacteurDeC5InduitQuiAtteintS(G)}
Let $G$ be a cubic graph which can be factored into $s(G)$ induced
odd cycles of length at least $5$, then $G$ is a permutation graph. Moreover, if $G$ has girth $5$ then $G$ is the Petersen graph.
\end{Thm}
\begin{Prf}
Let $\mathcal{F}$ be a $2$-factor of $s(G)$ cycles of length at least $5$ in
$G$, every cycle of $\mathcal{F}$ being an induced odd cycle of $G$. We
consider the \dm $\phi$ such that the edges of all cycles of
$\mathcal{F}$ are alternatively coloured   $\alpha$ and $\beta$
except for exactly one edge per cycle which is coloured with
$\delta$, all the remaining edges of $G$ being coloured $\gamma$. By
construction we have $B_{\phi}=C_{\phi}=\emptyset$ and
$A_{\phi}=\mathcal{F}$.

Let $xy$ be an edge connecting two distinct cycles of $\mathcal{F}$, say $C_1$ and $C_2$ ($x\in C_1$, $y\in C_2$). By Theorem \ref{Thm:Summarize}, since any edge in $C_1$ or $C_2$ can be coloured $\delta$,
we may assume that there is an edge in $C_1$, say $e_1$, adjacent to $x$ and coloured with $\delta$, similarly there is on $C_2$ an edge $e_2$ adjacent to $y$ and coloured with $\delta$. 
Let $z$ be the neighbour of $y$ on $C_2$ such that $e_2=yz$ and let $t$ be the neighbour of $z$ such that $zt$ is coloured with $\gamma$.
If $t\notin C_1$, there must be $C_3\neq C_1$ such that $t\in C_3$, by Theorem \ref{Thm:Summarize} again there is an edge $e_3$ of $C_3$, 
adjacent to $t$ and coloured with $\delta$. But now $\{e_1, e_2, e_3\}$ induces a subgraph with at least $5$ edges, a contradiction with Theorem \ref{Thm:Summarize}. 

It follows that $\mathcal{C}$ contains exactly two induced odd cycles and they are of equal length. Consequently $G$ is a permutation graph. When these cycles have length $5$, since $G$ has girth $5$ $G$ is obviously the Petersen graph.
\end{Prf}

\ignore{\textbf{Comments:} The index $s(G)$ used here is certainly not greater than
$o(G)$ the {\em oddness} of $G$ used by Huck and Kochol
\cite{HucKoc}. The oddness $o(G)$ is the minimum number of odd cycles in any
$2$-factor of a cubic graph (assuming that we consider graphs with
that property). Obviously $o(G)$ is an even number and  it is an
easy task to construct a cubic graph $G$ with $s(G)$ odd which
satisfies $0<s(G)<o(G)$. We can even construct
cyclically-$5$-edge-connected cubic graphs with that property with
$s(G)=k$ for any integer $k \neq 1$ (see Fouquet \cite{FouPhD} and Steffen \cite{Ste04}). It can be pointed out that, using a parity argument
(see \ignore{Lemma \ref{Lemma:APhiBPhiCPhi} or }\cite{Ste}) in a graph of oddness
at least $2$ the colour of minimum frequency is certainly used at least twice. In other words, $o(G)=2
\Leftrightarrow s(G)=2$.
}
When $G$ is a cubic bridgeless planar graph, we know from the Four Colour Theorem that $G$ is $3-$edge colourable and hence $\gamma(G)=1$. Albertson
and Haas \cite{AlbHaa} gave $\gamma(G) \geq
\frac{6}{7}-\frac{2}{35m}$ when $G$ is a planar bridgeless graph
with maximum degree $3$. Our Theorem \ref{Theorem:Gamma} improves
this lower bound (allowing moreover bridges). On the other hand,
they exhibit a family of planar  graphs with maximum degree $3$
(bridges are allowed) for which $\gamma(G)=\frac{8}{9}-\frac{2}{9n}$.

As Steffen in \cite{Ste04} we denote $g(\mathcal F)$ the minimum length of an odd cycle in a $2$-factor $\mathcal F$
 and $g^+(G)$ the maximum of these numbers over all $2$-factors. We suppose that $g^+(G)$ is defined, that is $G$ has at least
 one $2$-factor (when $G$ is a cubic bridgeless graphs this condition
 is obviously fulfilled).

When $G$ is cubic bridgeless, Steffen \cite{Ste04} showed that we
have~:
$$\gamma(G) \geq \quad {\rm max} \{1-\frac{2}{3g^+(G)},
\frac{11}{12}\}$$ The difficult part being to show that $\gamma(G)
\geq \frac{11}{12}$.

\begin{Thm} \label{Theorem:SteffenExtended} Let $G$ be a graph with maximum degree $3$.
Then $\gamma(G) \geq 1-\frac{2n}{(3n-|V_2|)g^+(G)}$.
\end{Thm}
\begin{Prf}
By Lemma \ref{Lemma:NoPendantVertex}, we may assume  $V_1 = \emptyset$. Hence, $m=\frac{1}{2}(2|V_2|+3|V_3|)$, moreover $n=|V_2|+|V_3|$, henceforth $m=\frac{3n-|V_2|}{2}$. 
We have $ \gamma(G)= 1-\frac{s(G)}{m}$,
obviously, $s(G) \leq \frac{n}{g^+(G)}$. The result follows.
\end{Prf}

\begin{Thm} \label{Theorem:maxdegree3_11_12}Let $G$  be a graph with maximum degree $3$ having at least one
$2$-factor. Assume that  $|V_2| \leq \frac{n}{3}$ and $g^+(G) \geq 11$ then
$\gamma(G) \geq \quad {\rm max} \{1-\frac{3}{4g^+(G)},
\frac{11}{12}\}$.
\end{Thm}

\begin{Prf}
By assumption we have $V_1 = \emptyset$. From Theorem
\ref{Theorem:SteffenExtended} we have just to prove that $\gamma(G)
\geq \frac{11}{12}$. Following the proof of Theorem
\ref{Theorem:Gamma}, we try to associate a set $P_e$ of at least $8$
distinct vertices of $V_3$ for each edge $e \in E_{\phi}(\delta)$ in
a \dm $\phi$ of $G$ such that
\begin{equation} \label{Equation:PrivateV3bis}
\forall e,e' \in E_{\phi}(\delta) \ \ P_e \cap P_{e'} = \emptyset
\end{equation}

Indeed, let $\mathcal F$ be  a $2$-factor of $G$ where each odd cycle
has length at least $11$ and let $C_1, C_2 \ldots C_{2k}$ be its set
of odd cycles. We have, obviously $s(G)\leq 2k$.  Let $V'_3$ and $V'_2$ be the
sets of vertices of degree $3$ and $2$ respectively contained in these
odd cycles . As soon as $|V'_3 |\geq 8s(G)$ we have
\begin{equation}\label{Equation:V3sur8}
   \gamma(G)= 1-\frac{s(G)}{m}=1-\frac{s(G)}{\frac{3}{2}|V_3|+|V_2|} \geq 1-\frac{\frac{|V'_3|}{8}}{\frac{3}{2}|V_3|+|V_2|}
\end{equation}

which leads to
$$\gamma(G) \geq 1-\frac{\frac{2|V'_3|}{24|V_3|}}{1+\frac{2}{3}\frac{|V_2|}{|V_3|}}$$ Since $|V_3|
\geq |V'_3|$, we have
$$\gamma(G) \geq 1-\frac{\frac{2}{24}}{1+\frac{2}{3}\frac{|V_2|}{|V_3|} } $$
and $$\gamma(G) \geq \frac{11}{12}$$ as claimed.

It remains the case where $|V'_3 |< 8s(G)$. Since each
odd cycle has at least $11$ vertices we have  $|V'_2|>11\times 2k - |V'_3 |>3s(G)$.
$$\gamma(G)= \frac{m-s(G)}{m}\geq \frac{m-\frac{|V'_2|}{3}}{m}$$
We have
$$\frac{m-\frac{|V'_2|}{3}}{m} \geq \frac{11}{12} $$ when
\begin{equation}\label{Equation:m4V2}
    m \geq 4|V'_2|
\end{equation}

Since $|V_2| \leq \frac{n}{3}$ we have $|V_3| \geq \frac{2n}{3}$ and
\begin{equation}\label{Equation:mm}
    m=3\frac{|V_3|}{2}+|V_2|=3\frac{n-|V_2|}{2}+|V_2|=3\frac{n}{2}-\frac{|V_2|}{2} \geq 4\frac{n}{3} \geq 4|V'_2|
\end{equation}
and the result holds.
\end{Prf}

{\bf Acknowledgment~:} the authors are grateful to the anonymous referees whose observations
 led to a number of improvments of this paper.

\bibliographystyle{plain}

\bibliography{Parcimonious}

\end{document}